\documentclass{raa}

\usepackage{epsfig}
\usepackage{multicol}
\usepackage{graphicx}
\usepackage{url}
\usepackage{natbib}
\usepackage{longtable}

\begin{document}
\title{Fixing the Reference Frame for PPMXL Proper Motions Using Extragalactic Sources}

   \volnopage{Vol.0 (200x) No.0, 000--000}      %%preserved for Editor. DOn't remove!
   \setcounter{page}{1}          %%starting page, preserved for Editor. DOn't remove!

\author{
Kathleen Grabowski
  \inst{1}
\and Jeffrey L. Carlin
  \inst{1}
\and Heidi Jo Newberg
  \inst{1}
\and Timothy C. Beers
  \inst{2} 
\and Li Chen
  \inst{3}
\and Licai Deng
  \inst{4}
\and Carl J. Grillmair
  \inst{5}
\and Puragra Guhathakurta
  \inst{6}
\and Jinliang Hou
  \inst{3}
\and S\'ebastien L\'epine
  \inst{7} 
\and Chao Liu
  \inst{4}
\and Xiaowei Liu
  \inst{8}
\and A-Li Luo
  \inst{4}
\and Martin C. Smith
  \inst{3}
\and Brian Yanny
  \inst{9}
\and Haotong Zhang
  \inst{4}
\and Zheng Zheng
  \inst{10}
}

\institute{Department of Physics, Applied Physics and Astronomy, Rensselaer Polytechnic Institute, 110 8th Street, Troy, NY 12180, USA (carlij@rpi.edu; newbeh@rpi.edu) \\
\and 
Department of Physics and JINA: Joint Institute for Nuclear Astrophysics, University of Notre Dame, 225 Nieuwland Science Hall, Notre Dame, IN 46556, USA \\
\and 
Shanghai Astronomical Observatory, 80 Nandan Road, Shanghai 200030, China\\
\and
Key Lab of Optical Astronomy, National Astronomical Observatories, Chinese Academy of Sciences, Beijing 100012, China\\
\and
Spitzer Science Center, 1200 E. California Blvd., Pasadena, CA 91125, USA\\
\and
UCO/Lick Observatory, Department of Astronomy and Astrophysics, University of California, Santa Cruz, CA 95064, USA\\
\and
Department of Physics and Astronomy, Georgia State University, 25 Park Place, Suite 605, Atlanta, GA 30303, USA\\
\and
Kavli Institute for Astronomy and Astrophysics, Peking University, Beijing 100871, China; Department of Astronomy, Peking University, Beijing 100871, China\\
\and
Fermi National Accelerator Laboratory, P.O. Box 500, Batavia, IL 60510, USA\\
\and
Department of Physics and Astronomy, University of Utah, UT 84112, USA\\
}

   \date{Received~~2014 month day; accepted~~2014~~month day}

\abstract{
We quantify and correct systematic errors in PPMXL proper motions using extragalactic sources from the first two LAMOST data releases and the V\`eron-Cetty \& V\`eron Catalog of Quasars. Although the majority of the sources are from the V\`eron catalog, LAMOST makes important contributions in regions that are not well-sampled by previous catalogs, particularly at low Galactic latitudes and in the south Galactic cap. We show that quasars in PPMXL have measureable and significant proper motions, which reflect the systematic zero-point offsets present in the catalog. We confirm the global proper motion shifts seen by \citet{wmz11}, and additionally find smaller-scale fluctuations of the QSO-derived corrections to an absolute frame. We average the proper motions of 158,106 extragalactic objects in bins of $3\times3$ degrees and present a table of proper motion corrections.
}

   \authorrunning{K. Grabowski, J.~L. Carlin, \& H.~J. Newberg, et al. }            %author_head in even pages
   \titlerunning{Corrections to PPMXL Proper Motions with LAMOST}  % title_head in odd pages

   \maketitle

\section{Introduction}           %% first-level sections will be auto-capitalized

Recent large spectroscopic surveys of Milky Way stars such as SDSS/SEGUE (e.g., \citealt{yrn+09}), RAVE (e.g., \citealt{szs+06a, kgs+13}), and LAMOST/LEGUE (e.g., \citealt{czc+12, dnl+12, zzc+12}) provide vast databases of line-of-sight (i.e., radial) velocities (RVs) for the study of Galactic kinematics. These data prove much more valuable if they can be combined with the tangential component of velocity via proper motions to derive the full three-dimensional velocities of large numbers of individual stars. However, proper motions are difficult to measure, and even more difficult to calibrate to an absolute frame. Furthermore, because the tangential velocity derived from proper motions is proportional to the product of the star's distance and its proper motion, any uncertainties in the absolute proper motions contribute even larger errors in derived tangential velocities. It is thus vital to characterize and remove any systematic uncertainties in the proper motion zero points using extragalactic sources to define an absolute reference frame.

This paper is motivated by the desire for a proper motion catalog to complement the LAMOST/LEGUE spectroscopic survey, which will obtain spectra of at least $6-7$~million stars over a 5-year period. LAMOST will observe much of the northern sky above $\delta = -10^\circ$ (for details on the survey design and target selection, see \citealt{cln+12, chy+12, lyh+14, ycl+12, zcy+12}), including nearly complete magnitude-limited samples to $r \approx 17$ over much of the high Galactic latitude regions available to the telescope. While much of the LAMOST footprint overlaps the regions surveyed by SDSS, there is also a large fraction of LAMOST spectroscopy in regions not observed by SDSS (especially at low Galactic latitudes near the anticenter, and in the south Galactic hemisphere). In order to fully exploit the LAMOST data for studies of Galactic kinematics, it is important to obtain well-calibrated proper motions that span the entire sky covered by this survey, and to comparable photometric depth. The additional advantage of LAMOST is that a survey observing such large numbers of objects will inevitably discover huge numbers of spectroscopically identified QSOs and galaxies (e.g., \citealt{wcj+10, wjc+10, hlx+13}). These extragalactic reference objects can then be used to test and/or calibrate the zero points of existing proper motion catalogs.

The UCAC4 catalog \citep{zfg+13} covers the entire sky and includes proper motions for over 105 million objects, with formal errors of 1-10 mas~yr$^{-1}$ per star, and systematics estimated to be $\sim1-4$~mas~yr$^{-1}$. However, its magnitude limit of $R=16$ is not deep enough to provide useful proper motions for all LAMOST stars. Proper motions from the combination of SDSS and USNO-B positions \citep{mml+04,mml+08} reach considerable depth, at high precision, and with ample extragalactic reference objects for absolute zero points. However, this is limited to the footprint of the SDSS photometric survey, and thus does not include much of the LAMOST survey area (especially at low Galactic latitudes near the anticenter). Thus, while proper motions given in SDSS are on a well-determined reference frame (given the vast numbers of quasars available in SDSS; see \citealt{dgk+11}), they are not suitable to provide a uniform catalog for the LAMOST survey.

A proper motion catalog that is well-matched to our needs for LAMOST is PPMXL \citep{rds10}. PPMXL was built by combining USNO-B1 and 2MASS positions (and relative proper motions from USNO-B1) and placing them on the ICRS via the PPMX \citep{rss+08} proper motion catalog. The resulting catalog consists of over 900 million objects covering the entire sky (though the quality of proper motions diminishes greatly in the southern celestial hemisphere), complete to a magnitude $V\sim20$.

The depth of PPMXL is also sufficiently faint to sample many extragalactic sources, whose numbers increase rapidly with magnitude. This is vitally important, as the PPMXL catalog is known to have significant systematic errors in its proper motions (e.g., \citealt{wmz11}). It is thus the goal of this paper to derive corrections to the PPMXL proper motion zero points in order to improve the utility of its stellar proper motions. This is done by identifying a large sample of extragalactic reference objects (including QSOs and galaxies), matching them to PPMXL, and then, using the fact that these objects should have zero proper motion, derive corrections as a function of position that improve the zero points of the PPMXL catalog.

\section{Data Selection of Extragalactic Objects}

\begin{figure}[!t]
\includegraphics[width=.95\linewidth]{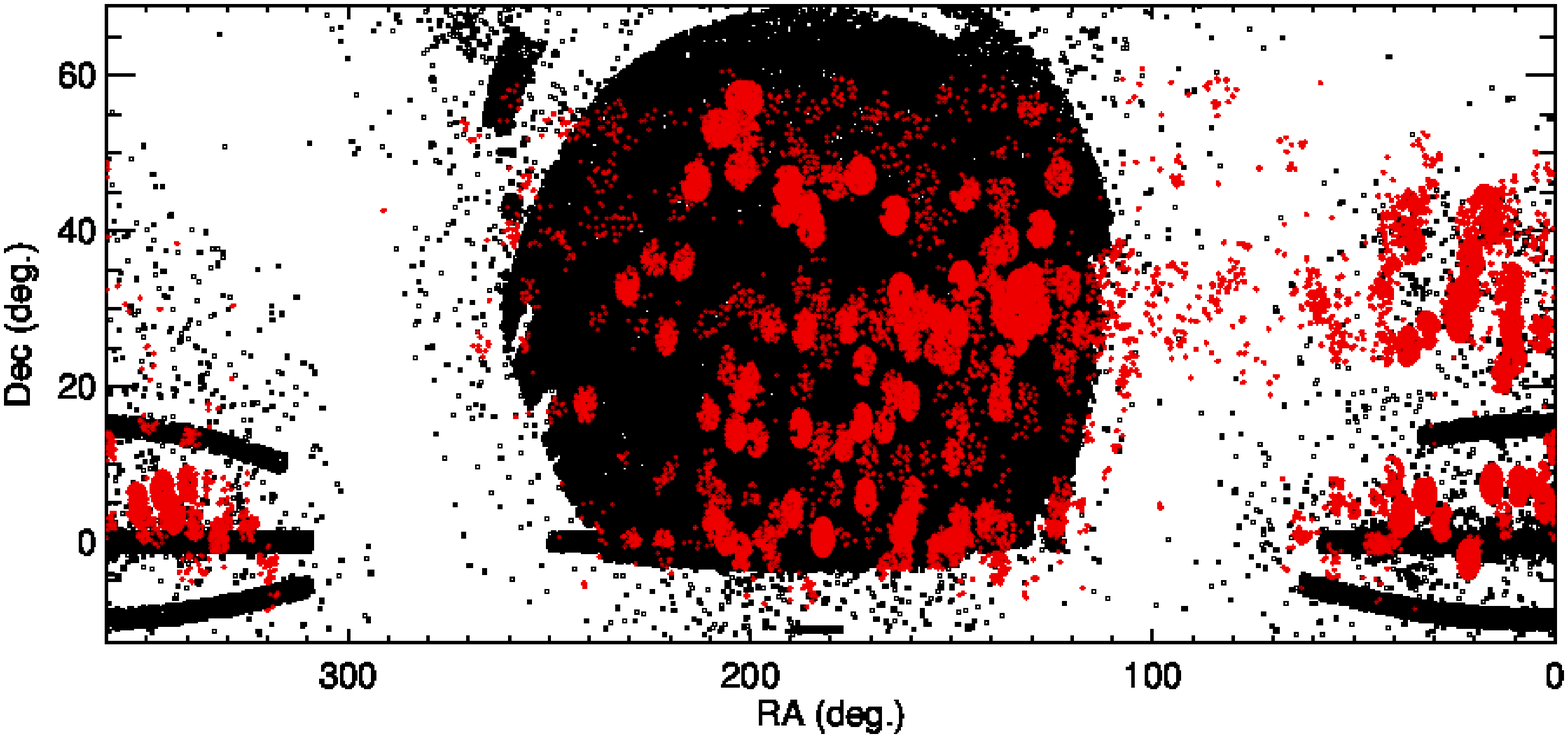}\\
\includegraphics[width=.95\linewidth]{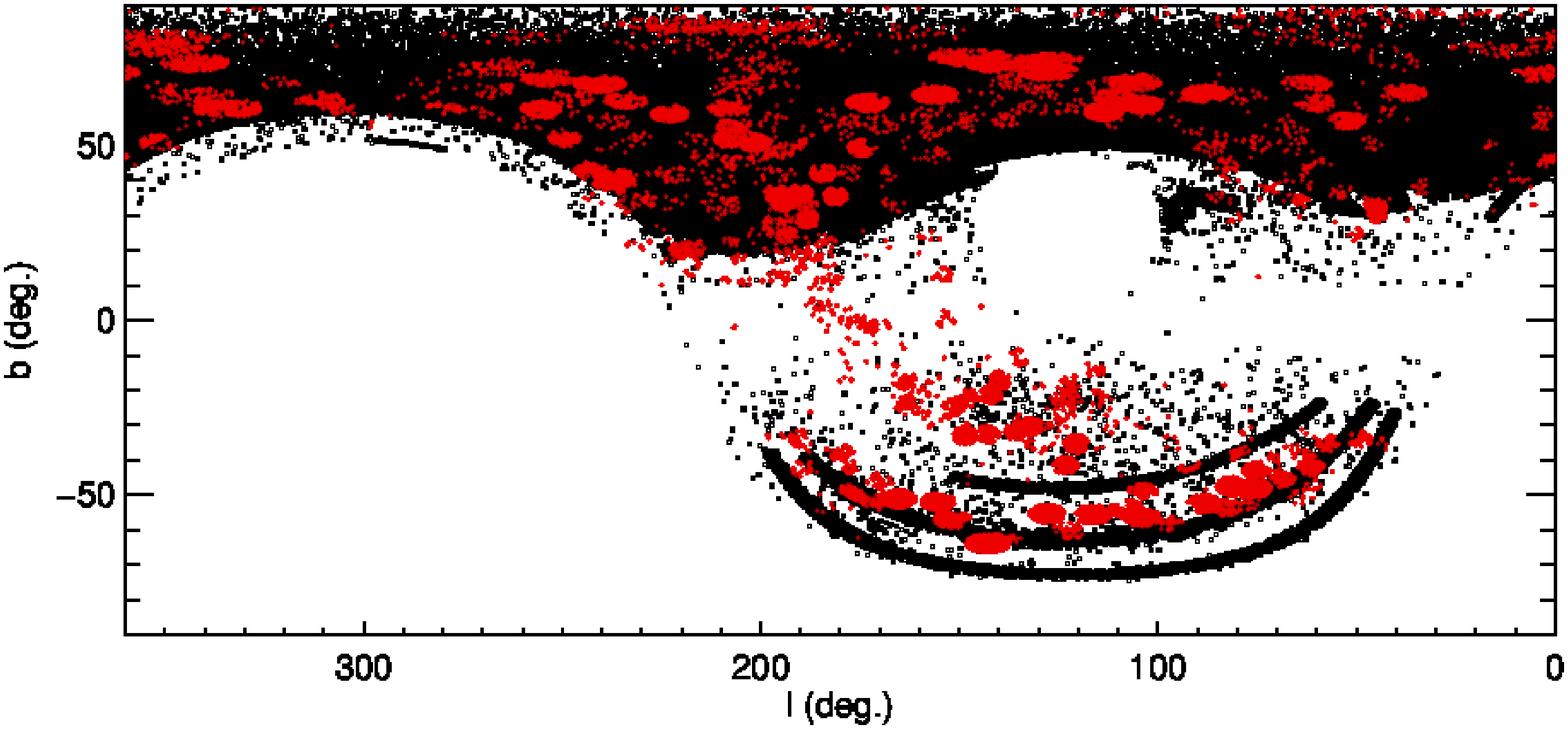}
\caption{Positions of objects in our extragalactic reference catalog. Spectroscopically identified QSOs and galaxies from LAMOST are denoted by red points, while V{\`e}ron quasars are denoted by black points. The upper panel shows the location of objects in equatorial coordinates and the lower panel shows their location in Galactic coordinates. The LAMOST objects improve the sky coverage mostly at low Galactic latitudes and in the south Galactic cap, where SDSS data are sparse.}
\label{coords}
\end{figure}

We began with a list of QSOs from the 13th edition of the V{\`e}ron-Cetty \& V{\`e}ron Quasar Catalog \citep{vv10}. We identified matches in PPMXL for a total of 125,276 QSOs from this catalog, which is slightly higher than the number studied by \citet{wmz11} of 117,053 quasars. While this is a vast data set with which to correct the PPMXL proper motions, the majority of the QSOs in this catalog come from SDSS, and are thus limited to the SDSS footprint on the sky. We thus supplemented this QSO catalog with LAMOST data from the Pilot Survey \citep{lzz+12} and internal Data Releases 1 and 2 (DR1 and DR2). Objects classified as ``QSO" or ``GALAXY" by the LAMOST pipeline were matched to PPMXL within a radius of 1 arcsecond. Galaxies
were included to supplement the paucity of quasars identified, although galaxy proper motions are more difficult to measure accurately because their morphologies are less point-like than stars and QSOs. We visually examined a few dozen of the spectra covering a range of signal-to-noise to confirm that their identification as QSOs or galaxies is reliable. This inspection confirms that only a small fraction ($<20\%$) of the objects are misidentified, and most of these have too little signal to be classified at all. Of the 3,747 QSOs at Galactic latitudes $b>30^\circ$ in LAMOST DR1 (i.e., in the region overlapping the SDSS northern footprint), 42\% have matches within 1~arcsecond in the V{\`e}ron-Cetty \& V{\`e}ron catalog. The 58\% that do not match are mostly at the faint end of the LAMOST targets, and have low S/N. Nevertheless, we believe that most of these are legitimate newly-identified QSOs. There are 4,750 quasars and 10,892 galaxies in DR1 (including Pilot Survey data), and an additional 3,130 QSOs and 14,058 galaxies in DR2 (as of April 2014). In total, our catalog of extragalactic sources contains 158,106 objects, of which 24,950 are galaxies and 133,156 are quasars.

Figure~\ref{coords} shows the location of QSOs and galaxies spectroscopically identified by LAMOST, in addition to the location of quasars from \citet{vv10}. While the V{\`e}ron-Cetty \& V{\`e}ron Quasar Catalog covers much of the northern Galactic cap, there is very little data near the plane of the Galaxy and in the southern Galactic cap. LAMOST has begun to fill in the Galactic plane near the anti-center and the southern Galactic cap for objects with a declination $>-15^\circ$.

\section{Correcting Proper Motions}

The proper motions of extragalactic sources should be zero, so the proper motion reported in PPMXL reflects the combination of random and systematic errors in this catalog. To correct proper motions for individual stars, the systematic errors should be subtracted from the values in PPMXL. 
To measure the zero-point errors, we average the proper motions of extragalactic objects in binned regions on the sky.\footnote{To keep the calculation and application of offsets simple, we bin in ($\alpha, \delta$) celestial coordinates rather than sky area as viewed from Earth ($\alpha \cos{\delta}, \delta$). This decreases the area sampled at high declinations, thus potentially reducing the number of objects included in these bins. However, given the limitations of LAMOST to $-10 < \delta < 65^\circ$, this effect reduces the area by at most a factor of $\sim1/2$.}
The size of the bins was adjusted so that at least three extragalactic objects would be included in each calculation, and yet provide as high a resolution as possible.
If the bins are too small, the errors in the corrections may be dominated by the 4-10~mas~yr$^{-1}$ random errors for individual objects in PPMXL, and thus be larger than the typical zero-point corrections. 
Our requirement of at least three objects in each bin allowed us to reject objects with absurdly high proper motions before averaging. 
The binning also effectively ``smooths'' across gaps where no data may be available in a particular small region. 
To calculate the average, we used an iterative 3-$\sigma$ clipping algorithm over objects binned together over 3x3 square degree boxes in RA and Dec. Bins were only included if they had at least 3 objects within the bin. 

Published papers measuring similar proper motion corrections include bin sizes ranging from one square degree, as used by \citet[though a $3\times3$ bin moving average was used, so the effective resolution is larger than a degree]{rds10},
to $15\times15$ degree bins as used by \citet{l14}. When we averaged objects in $15\times15$ degree bins similarly to \citet{l14}, it was more difficult to see both large-scale trends and smaller variations of QSO proper motions with RA and declination than with smaller bin sizes. \citet{l14} looked at proper motion trends in Galactic coordinates, but we have chosen to derive corrections in equatorial coordinates. We chose this approach because most zonal errors in proper motions are likely to be contributed by observational effects (e.g., differential refraction) that will depend more directly on the position in the sky as observed at the telescope, and are thus more likely to be correlated with equatorial coordinates.
We continue with analysis from $3\times3$ degree bins because the mean proper motions are similar to those from the single degree bins. This implies that the proper motion zero-point shifts do not vary that much on small scales.

\begin{figure}[!t]
\includegraphics[width=0.9\linewidth]{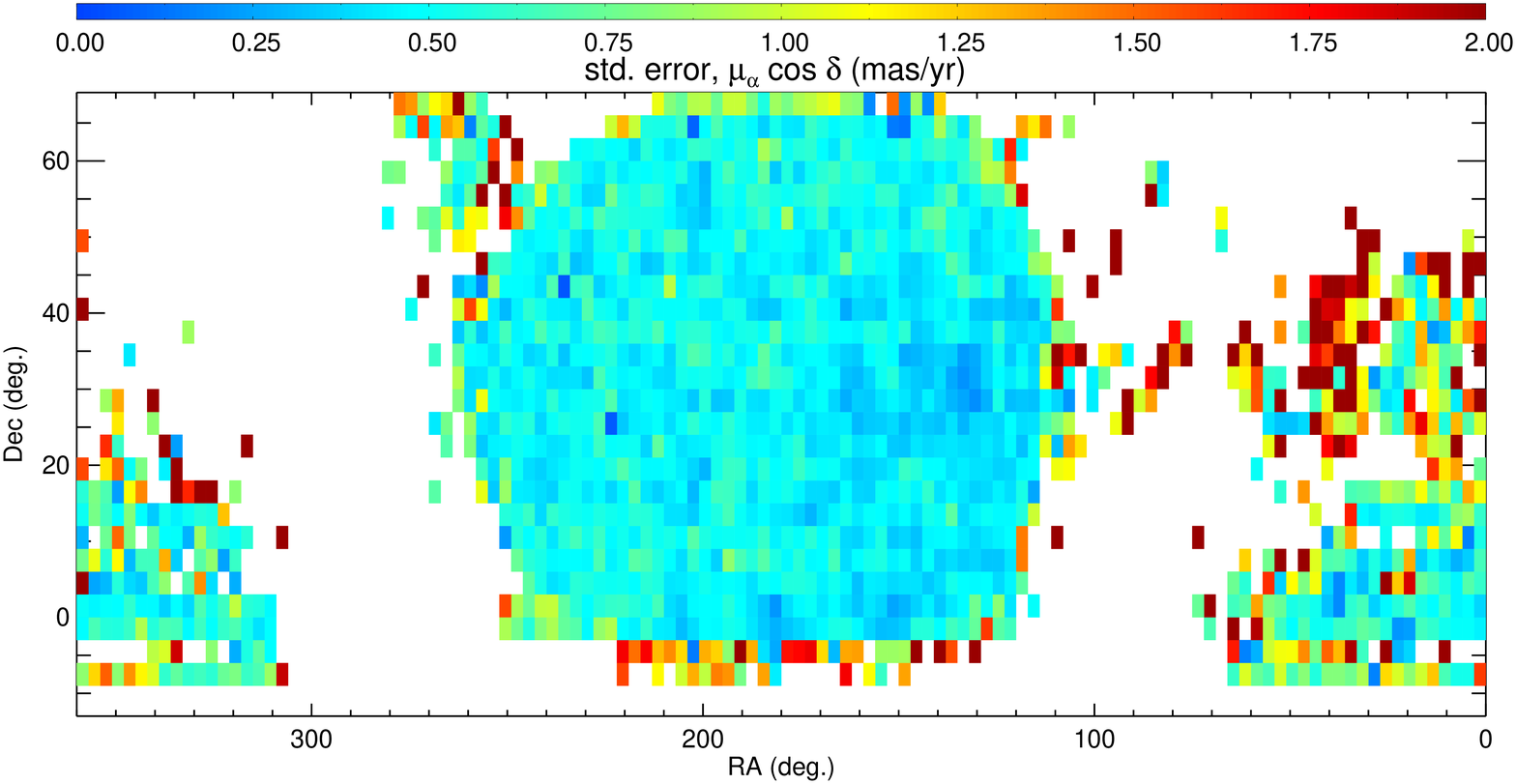}\\
\includegraphics[width=0.9\linewidth]{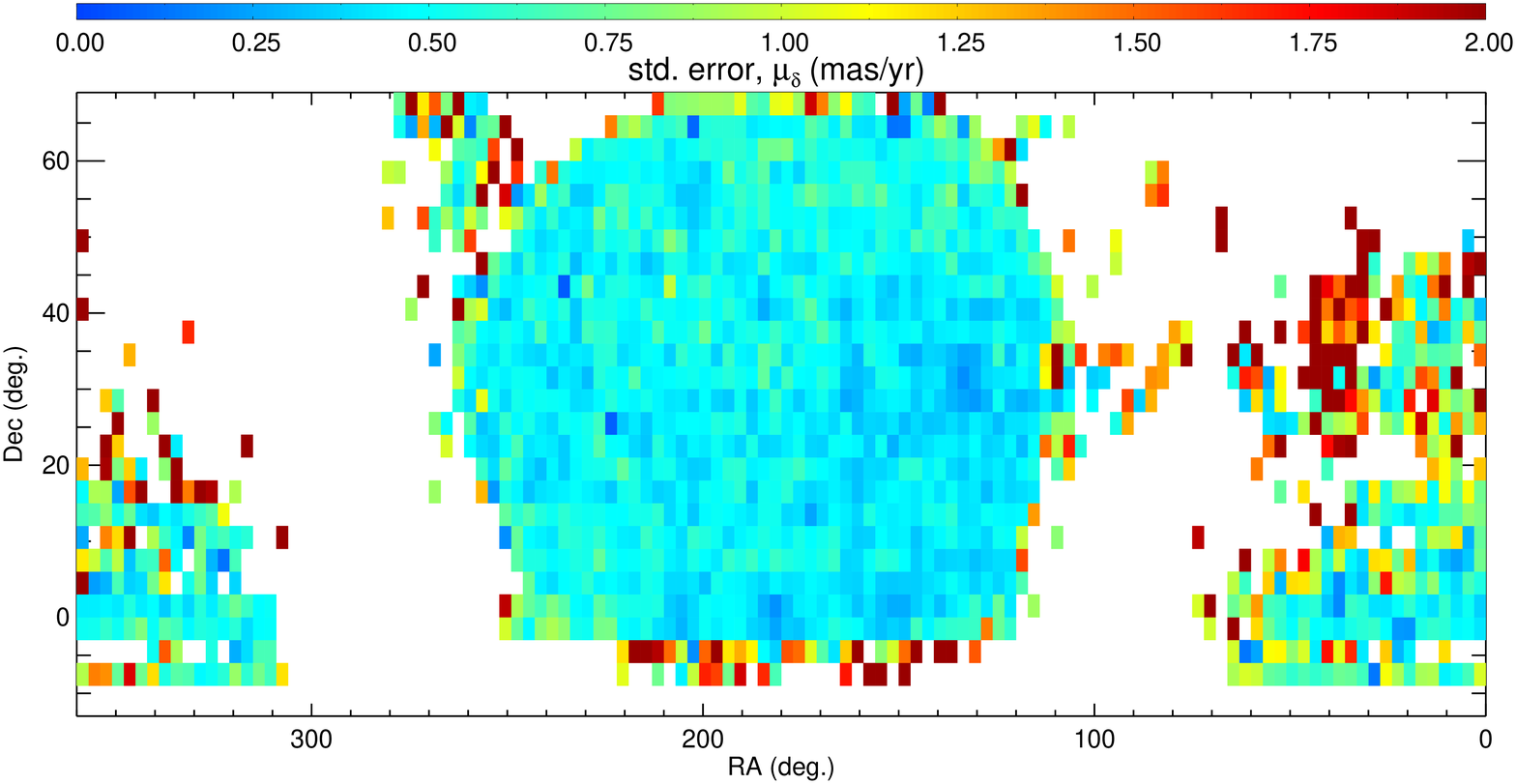}
\caption{Standard error on the sigma-clipped mean proper motions for extragalactic objects in $3\times3^\circ$ bins as a function of position (in equatorial coordinates). The top panel shows the standard error on the mean $\mu_{\alpha} \cos{\delta}$, and the lower panel shows the $\mu_{\delta}$ errors, both color-coded in units of mas~yr$^{-1}$.}
\label{pmstderr}
\end{figure}

Corrections made with the $3^\circ$ values will be effective for individual objects, and their mean values are determined much more reliably than in the $1^\circ$ bins due to the larger number of objects in each bin. There are at least 100 objects in 25\% of the $3\times3$ degree bins, while no single-degree bin has more than 102 objects. The standard error on the sigma-clipped mean proper motion for each $3\times3^\circ$ bin is shown in Figure~\ref{pmstderr} as a function of position on the sky. The color coding depicts the uncertainty on the mean proper motion correction in each bin. Because the standard error decreases as $n^{-1/2}$, the values in this figure are proportional to the number of objects in each bin. The errors are small over the large contiguous areas that contain SDSS data, and are larger in the regions between SDSS coverage that are being filled in by LAMOST. These measurements will improve as more LAMOST data are added to the sample.

\begin{figure}[!t]
\includegraphics[width=0.9\linewidth]{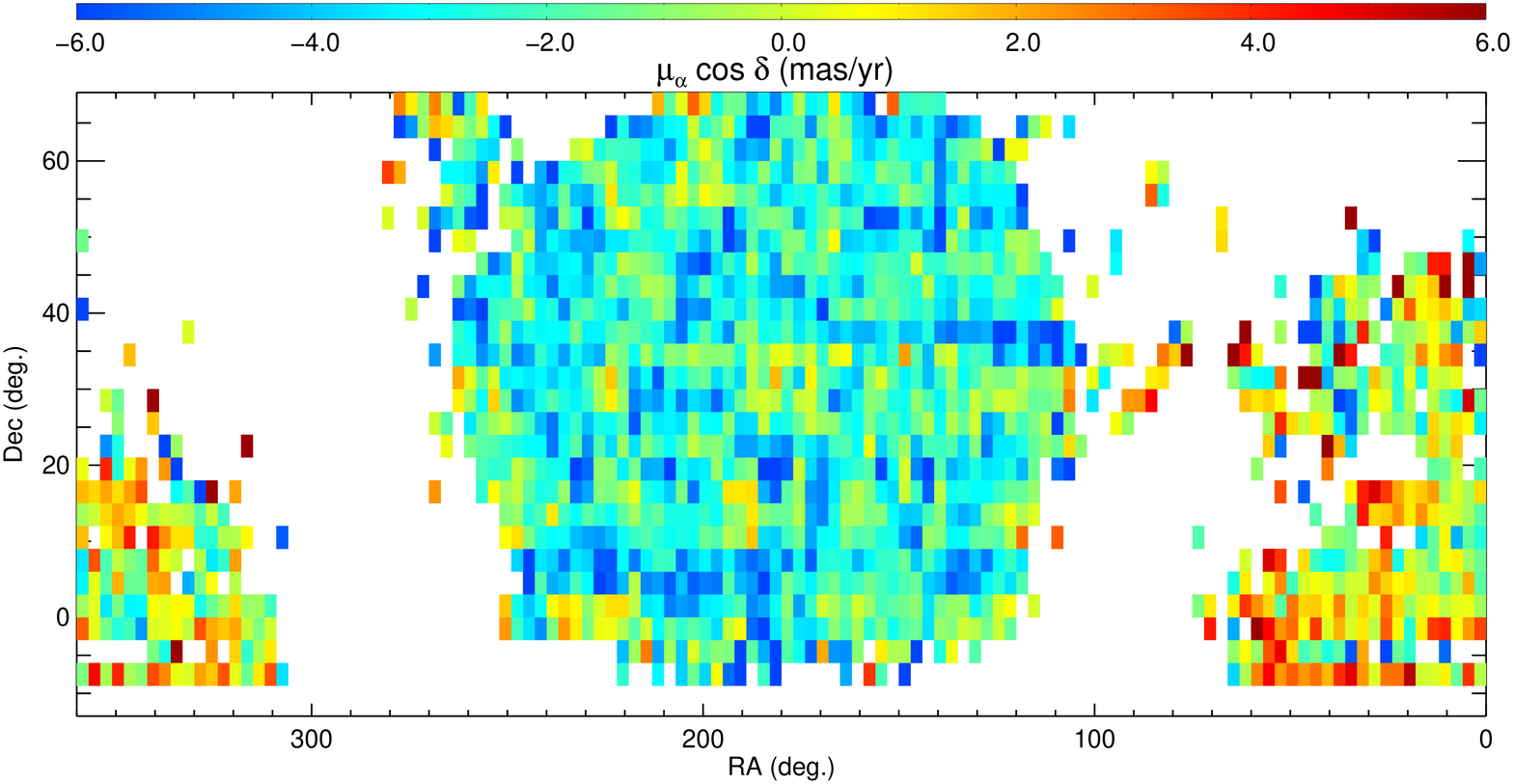}\\
\includegraphics[width=0.9\linewidth]{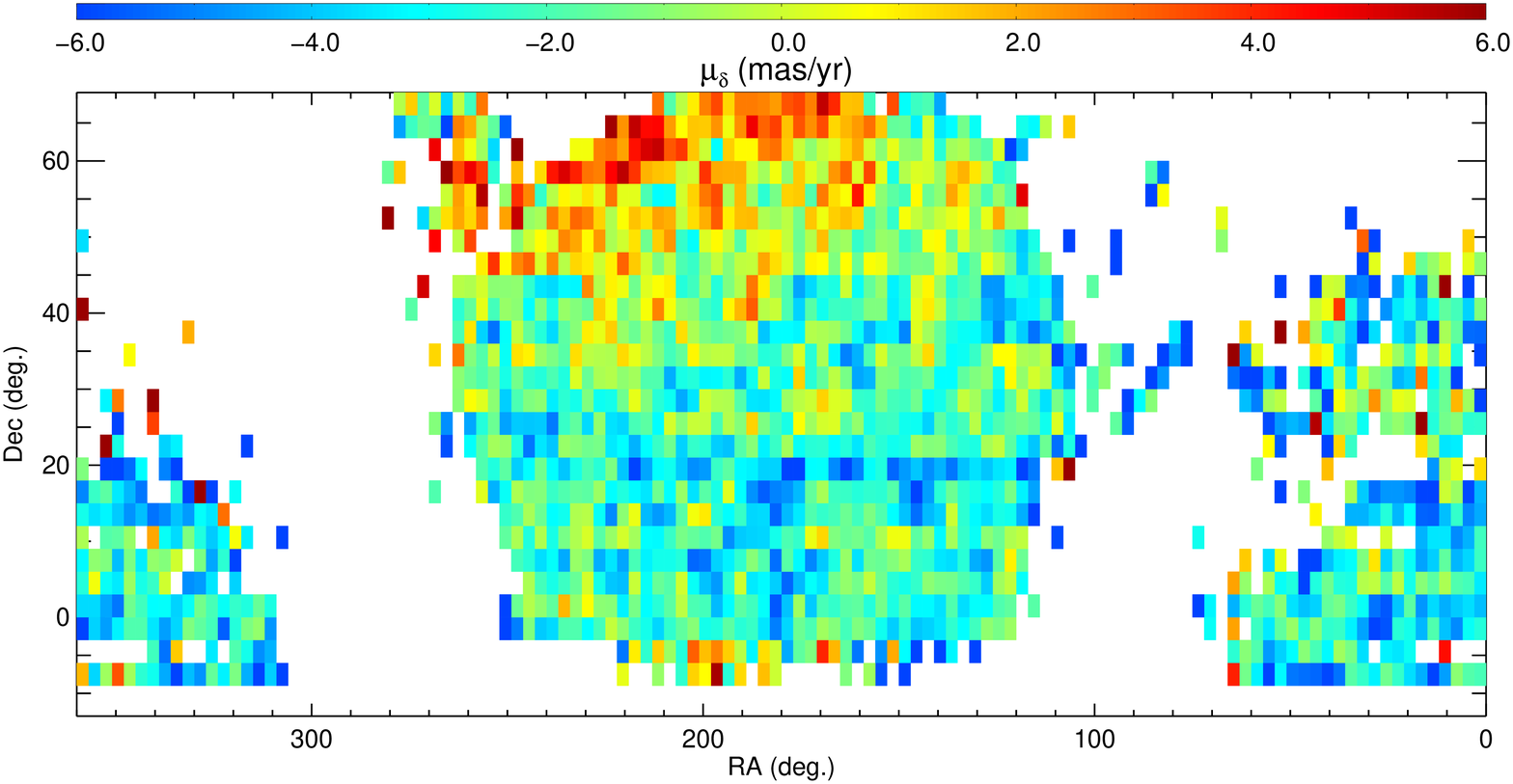}
\caption{Sigma-clipped mean proper motions for extragalactic objects in $3\times3^\circ$ bins as a function of position (in equatorial coordinates). The top panel shows $\mu_{\alpha} \cos{\delta}$, and the lower panel shows $\mu_{\delta}$, both color-coded in units of mas~yr$^{-1}$.}
\label{pm}
\end{figure}

In Figure~\ref{pm}, the sigma-clipped mean proper motion in Right Ascension and Declination is shown (encoded by color) for each 3-degree bin on the sky (in equatorial coordinates).  In the upper panel, there are small-scale variations between neighboring bins, and in the area covering the southern Galactic cap (RA$ > 300^\circ$ and RA$ < 70^\circ$), proper motion as a function of RA increases from a negative number to high proper motion values as the declination decreases. The lower panel shows a clear gradient of $\mu_\delta$, with the higher declinations showing a higher proper motion. There are also small-scale variations over the sky in $\mu_\delta$, even across $3\times3$ degrees. Furthermore, comparison of Figure~\ref{pm} with Figure~\ref{pmstderr} shows that these fluctuations on small scales are not due to corresponding uncertainties in the measurements, and are thus real variations in the PPMXL catalog. Fitting a smooth function to these residuals would not take all of the small-scale variations into account.

Figure~\ref{pms_mc} shows histograms of the proper motions of all objects in our catalog before (dashed black histogram) and after (solid black curve) applying the corrections we have derived. We fit Gaussians to the corrected proper motions in each panel, which we show as solid blue curves. These Gaussians are centered at -0.13 mas~yr$^{-1}$ and -0.28 mas~yr$^{-1}$ in $\mu_\alpha \cos{\delta}$ and $\mu_\delta$, respectively, with $\sigma = 4.34$ and $4.08$~mas~yr$^{-1}$. The scatter in the proper motions after correcting for systematic shifts is thus consistent with (though slightly smaller than) the $\sim5$~mas~yr$^{-1}$ median formal errors on the individual proper motions given in PPMXL. To assess the reliability of the PPMXL random errors, we turn to a Monte Carlo approach. For each object in Figure~\ref{pms_mc}, we calculate a realization of the proper motion using: $\mu_{\rm obs} = \mu_{\rm true} + randomn*\delta\mu$, where $\delta\mu$ is the uncertainty in the catalog, $\mu_{\rm true}$ is the ``true'' proper motion (in this case, zero), and randomn is a normally distributed random number. Each object is resampled 1000 times, and the resulting distribution of proper motions is overplotted in Figure~\ref{pms_mc} as a red dot-dashed histogram. The width of the resulting distribution is slightly larger ($\sim5.2$ mas~yr$^{-1}$ in both panels) than that of the corrected proper motions, suggesting that the PPMXL formal errors are in fact slightly overestimated by roughly 25\%. 
We also note the non-Gaussian tails to the distributions -- these could arise either due to truly large uncertainties in PPMXL, or by errors in the cross-matching of objects in the tables compiled for this work. The Monte Carlo test seems to confirm that these non-Gaussian tails are indeed a feature of the PPMXL data, which includes many objects whose proper motion errors are greater than 7~mas~yr$^{-1}$. Finally, we note that the fact that the corrected proper motions have a narrower distribution than found by the Monte Carlo procedure suggests that there are not significant smaller-scale (i.e., less than $3^\circ$) proper motion variations detectable in the current data set.

\section{Discussion}
One means of correcting the PPMXL proper motions is that of \citet{wmz11}, who fit separate sine curves to the proper motion residuals of QSOs in the RA and Dec directions. In this case, the dearth of data near the Galactic plane required extrapolation over large areas of blank sky. Furthermore, while it fit the data fairly well in \citet{wmz11}, the use of a sine function is arbitrary, and does not allow much freedom to fit smaller-scale spatial variations of the zero points. Another correction method was used by \citet{cdn+13}, who fit a second-order polynomial surface to proper motions of quasars as a function of position. This form better captures the global shape of the residuals, but cannot adjust to small-scale variations. Furthermore, the polynomial function fit by \citet{cdn+13} ranged from $0-360^\circ$ in RA, and thus is not continuous across $\alpha = 360^\circ = 0^\circ$ (note that these authors did not use data near $\alpha = 0^\circ$ in their study, so were not concerned by this issue). 

\begin{figure}[!t]
\begin{tabular}{c c}
\includegraphics[width=.49\linewidth]{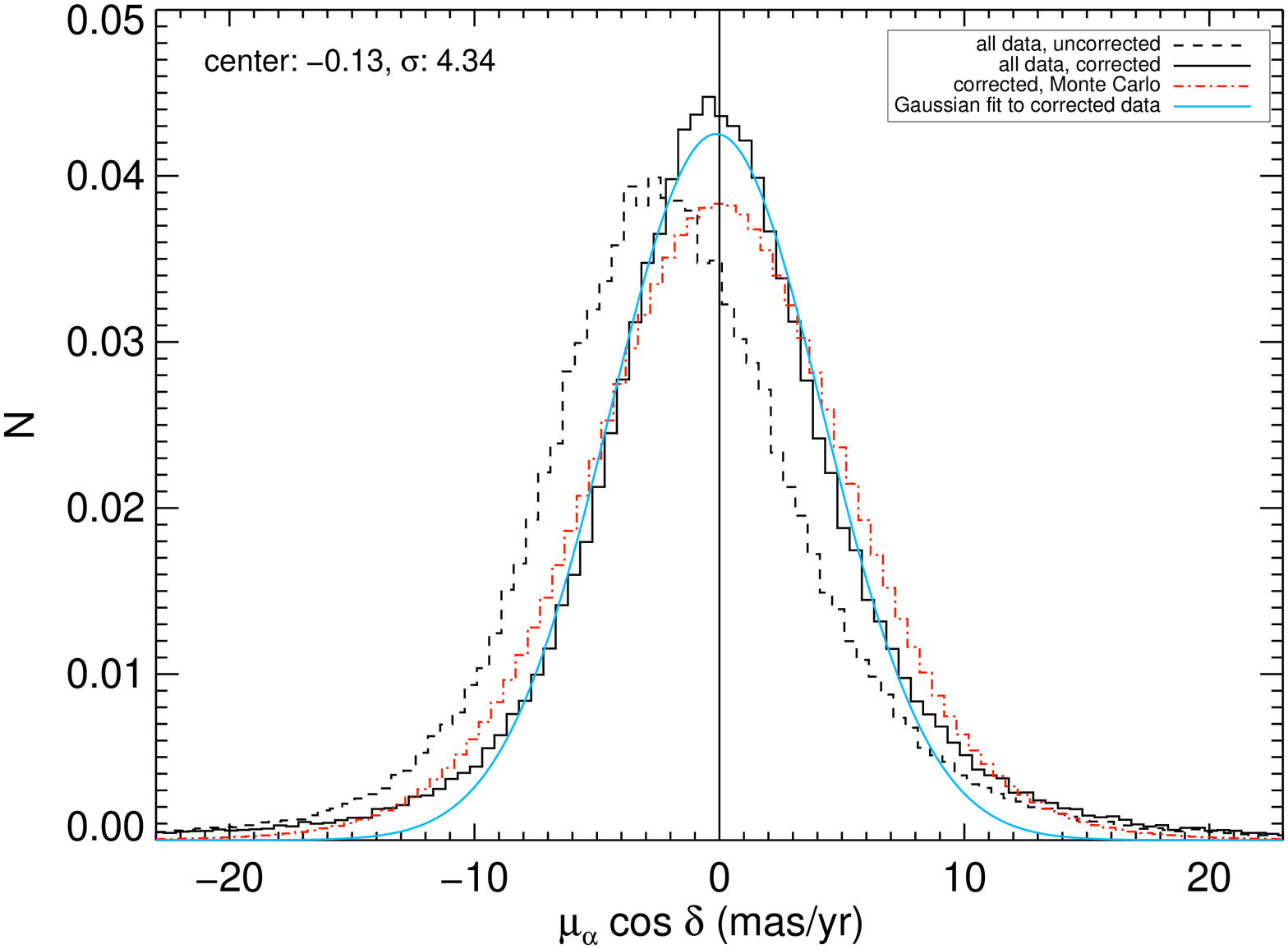}&
\includegraphics[width=.49\linewidth]{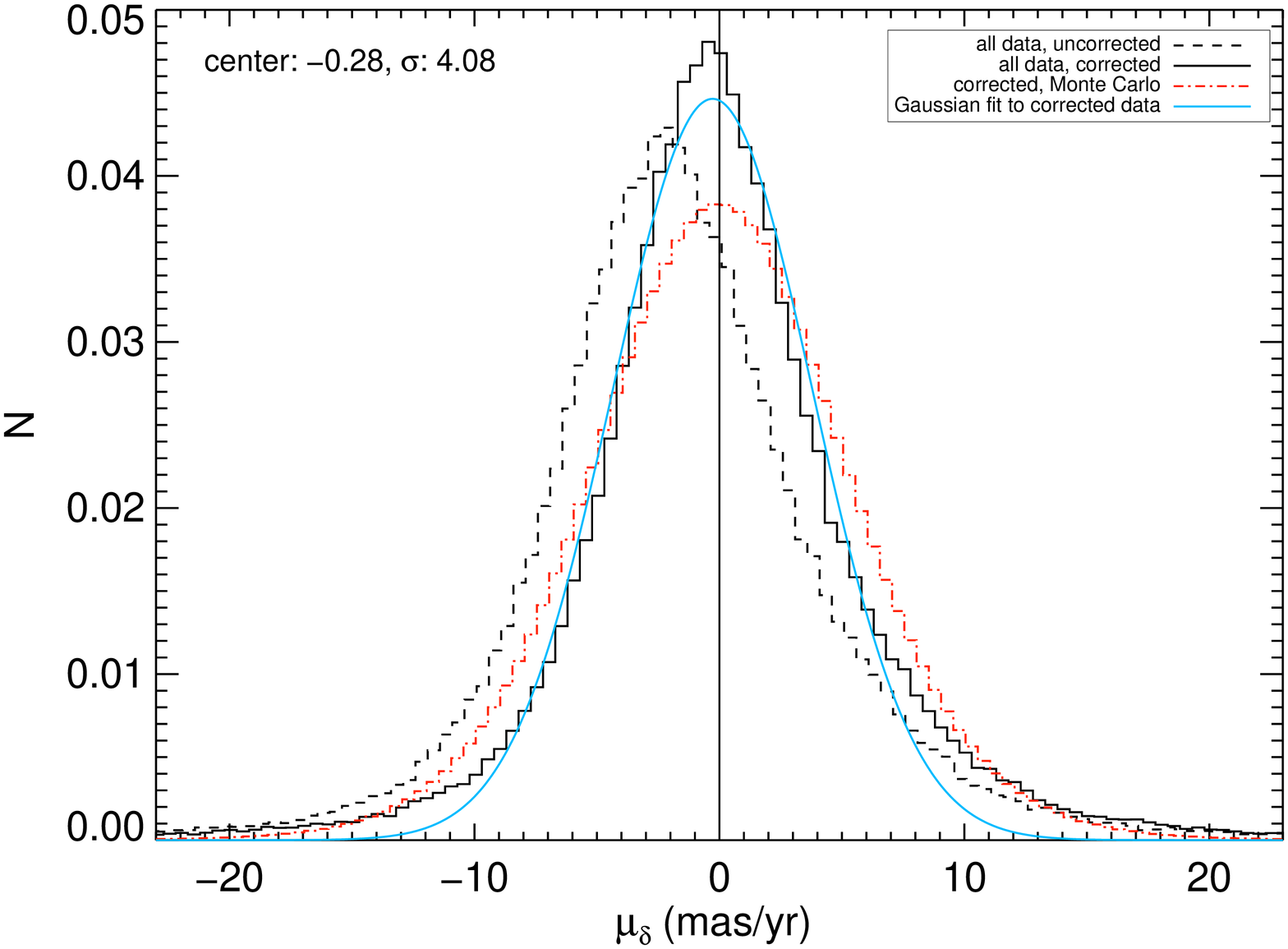}
\end{tabular}
\caption{Proper motions of all individual extragalactic objects from our input catalog. Left and right panels show $\mu_\alpha \cos{\delta}$ and $\mu_\delta$, respectively. The black dashed histogram depicts the ``raw'' proper motions before any corrections have been applied. The solid black histogram  shows the distribution after applying the corrections seen in Figure~\ref{pm} and given in Table~\ref{pmtable}. Gaussian fits to the distributions after correcting the zero points are solid blue curves, with the center and Gaussian $\sigma$ given in each panel. Finally, the red dot-dashed histogram shows the results of our Monte Carlo resampling from the formal PPMXL errors. The width of this curve matches that of the corrected data, suggesting that the PPMXL errors are reasonably estimated. Note also that the Monte Carlo curve reproduces the non-Gaussian tails in the data, again suggesting that the PPMXL errors are well characterized.
}
\label{pms_mc}
\end{figure}

\begin{figure}[!t]
\begin{tabular}{c c}
\includegraphics[width=0.49\linewidth]{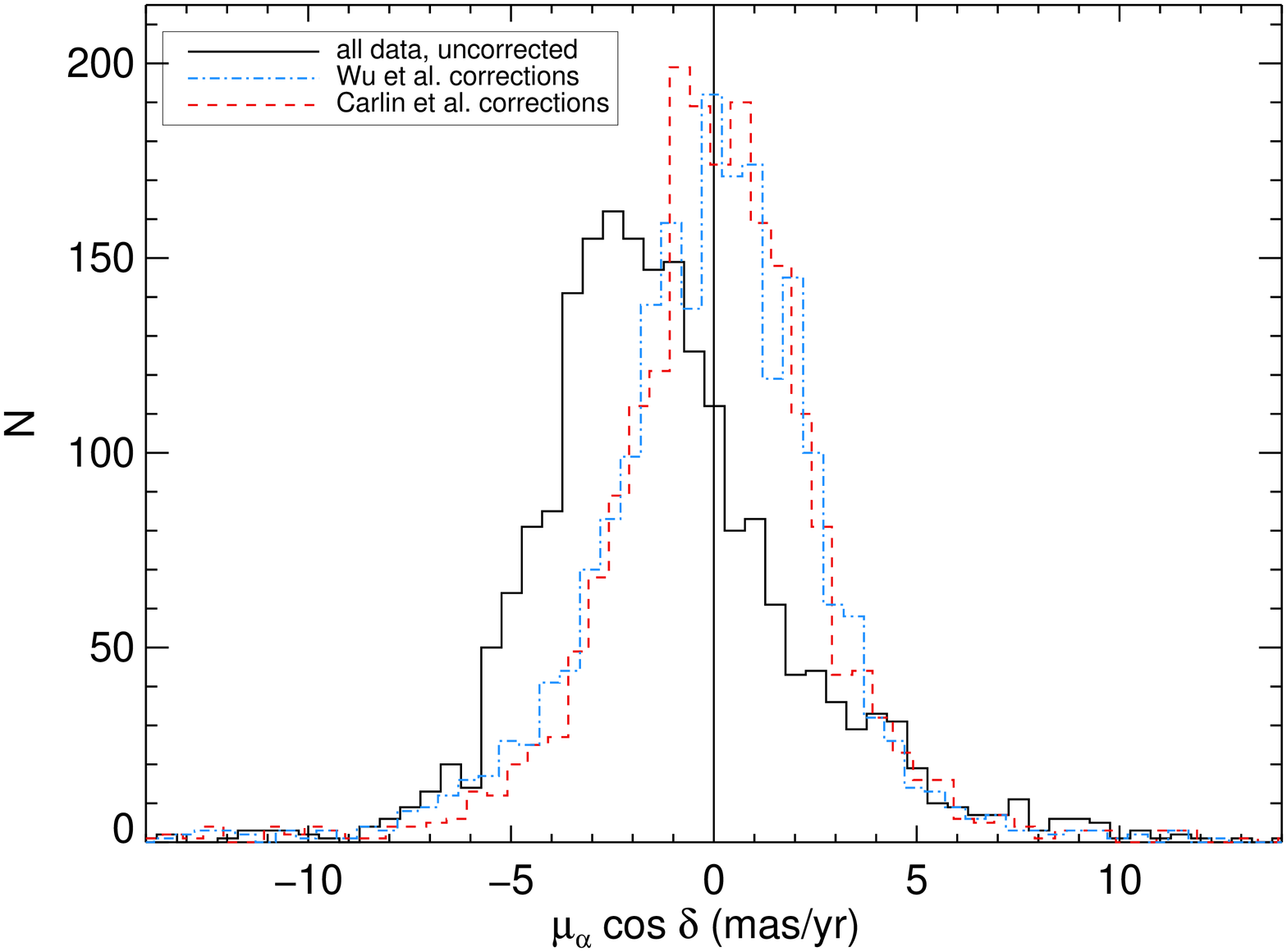}&
\includegraphics[width=0.49\linewidth]{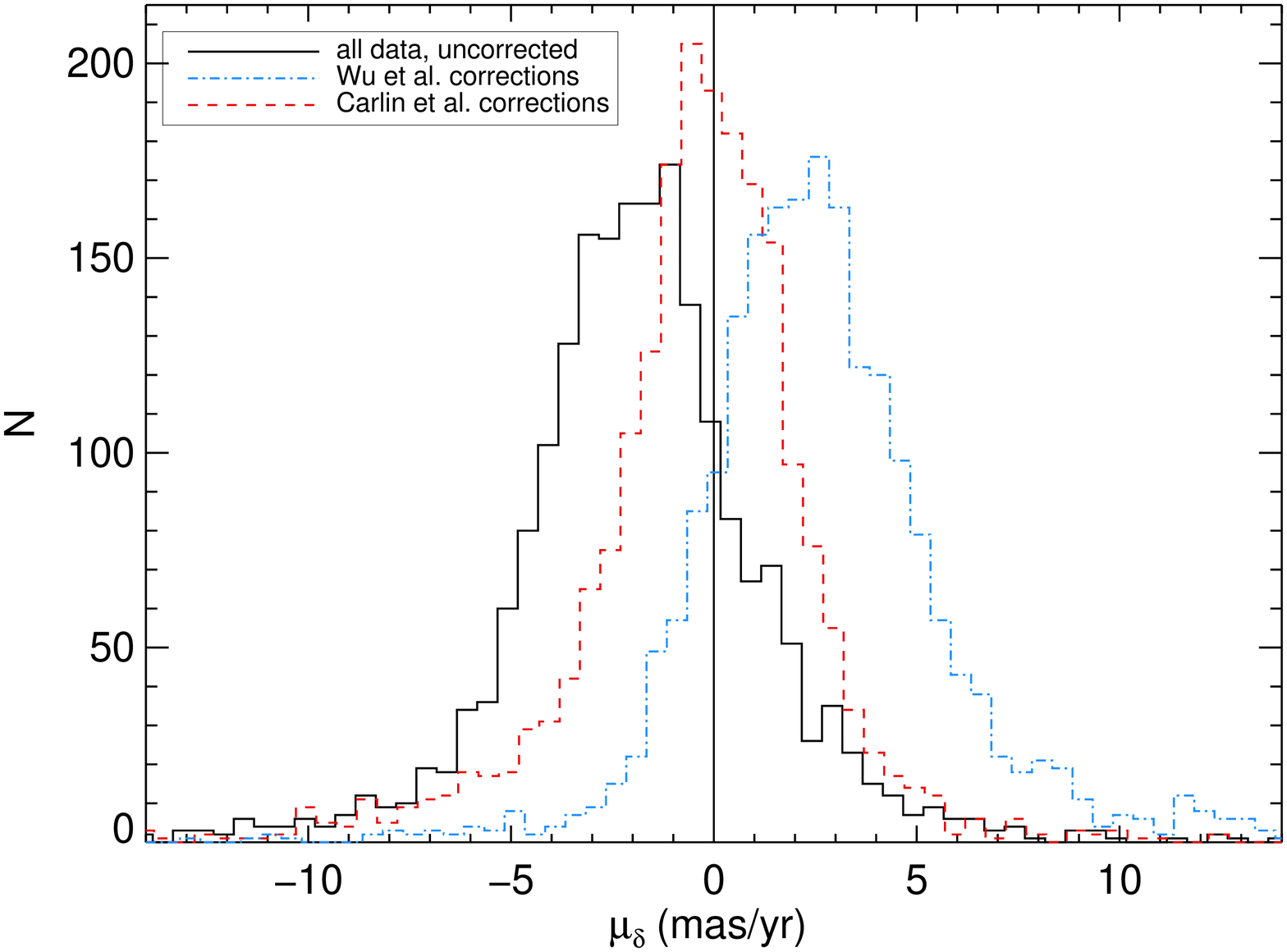}
\end{tabular}
\caption{Mean proper motions of all extragalactic objects in $3\times3^\circ$ bins, with $\mu_\alpha \cos{\delta}$ in the left panel and $\mu_\delta$ in the right panel (both in units of mas~yr$^{-1}$). 
In both panels, the solid black histograms show uncorrected mean proper motions of extragalactic objects from LAMOST and the V{\`e}ron-Cetty \& V{\`e}ron Quasar Catalog. Blue dot-dashed histograms are the results after applying the corrections of \citet{wmz11}, and red dashed histograms show mean proper motions after correcting via the method of \citet{cdn+13}. The Wu et al. and Carlin et al. methods produce nearly identical results in $\mu_\alpha \cos{\delta}$, while the systematic shift remaining in the Wu et al. corrected proper motions is obvious in the right panel. }
\label{compare}
\end{figure}

In Figure~\ref{compare}, we show histograms of sigma-clipped mean proper motions of uncorrected extragalactic objects in $3\times3^\circ$ bins as the solid black curves. Blue dot-dashed histograms show the mean proper motions in the same bins after correcting by the \citet{wmz11} method, and red dashed bins are the result after correcting via the polynomials of \citet{cdn+13}. 
The uncorrected data (solid black curves) show the systematic shifts pointed out by \citet{wmz11} for PPMXL proper motions in both dimensions. 
Based on the objects chosen from our compilation of LAMOST and \citet{vv10} data, the mean proper motions for all bins are ($\mu_{\alpha} \cos{\delta}$, $\mu_\delta$)=(-1.33, -1.69)~mas~yr$^{-1}$. 
The corrections suggested by \citet{wmz11} shift the extragalactic objects to ($\mu_{\alpha} \cos{\delta}$, $\mu_\delta$)=(0.02, 2.69)~mas~yr$^{-1}$. While the proper motion in RA is very close to zero, as expected for distant quasars, the proper motion in declination is over-corrected so that the mean of the corrected proper motion is greater than zero.\footnote{We note that ideally one would first fit the residuals in one dimension (say, RA), then apply this correction before fitting the other dimension (in this example, Dec). However, from their paper it appears that \citet{wmz11} independently fit corrections in RA and Dec separately, and did not correct one dimension before fitting the other. Thus, we attempted to mimic their procedure and applied the functions separately, which is likely the reason for the overcorrection in $\mu_\delta$ (i.e., the data have been ``double-corrected'' by this method).} The corrections derived by \citet{cdn+13} shift the LAMOST and V{\`e}ron objects to ($\mu_{\alpha} \cos{\delta}$, $\mu_\delta$)=(0.07, -0.29)~mas~yr$^{-1}$. The mean values of the Carlin et al. fits are very near zero in both dimensions, and have narrower distributions than the raw data ($\sim10-20\%$ smaller standard deviation). However, these fits are not capable of capturing small-scale (less than a few degrees) variations in the zero points.

\begin{figure}[!t]
\begin{tabular}{c c}
\includegraphics[width=.49\linewidth]{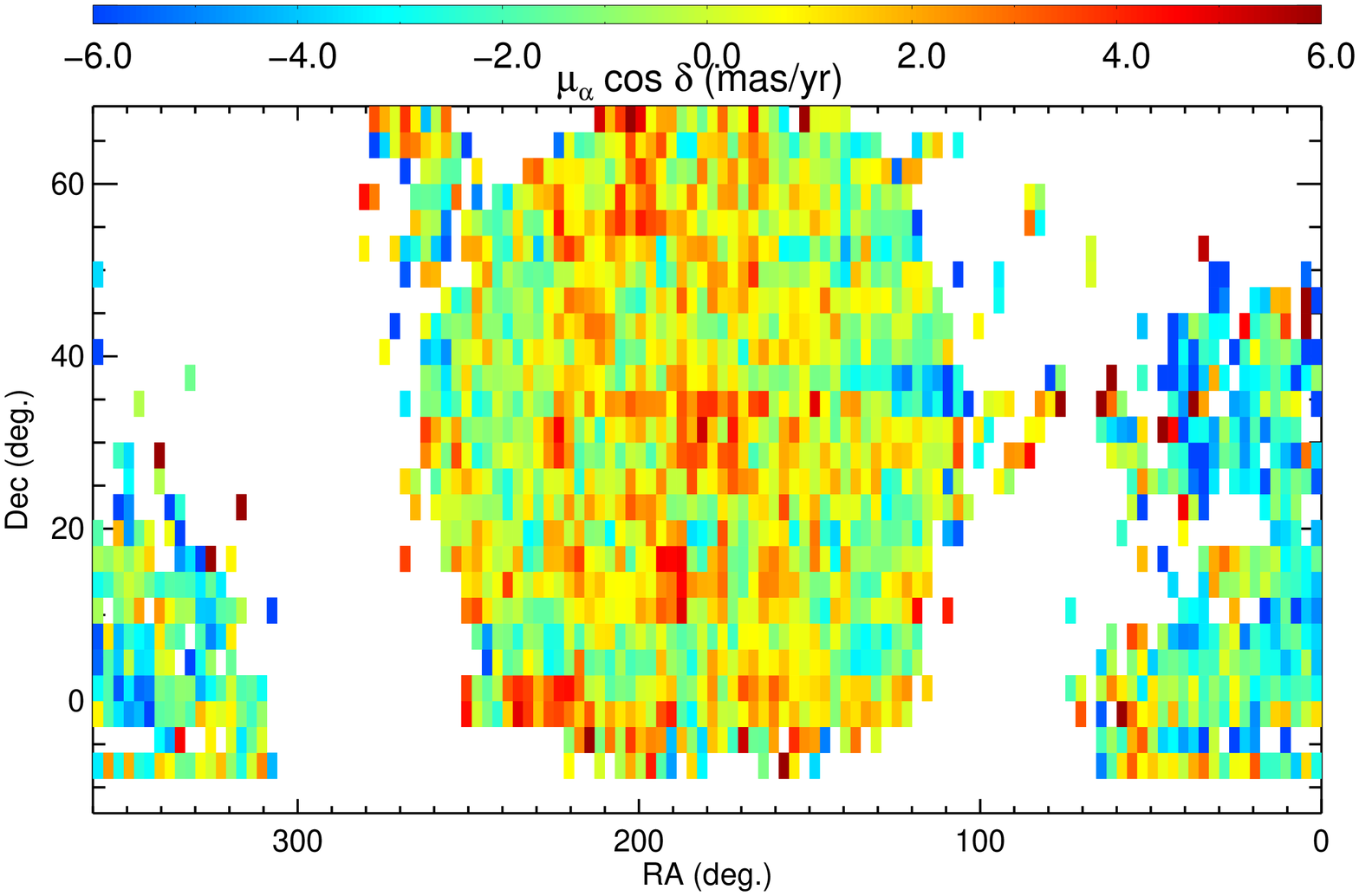}&
\includegraphics[width=.49\linewidth]{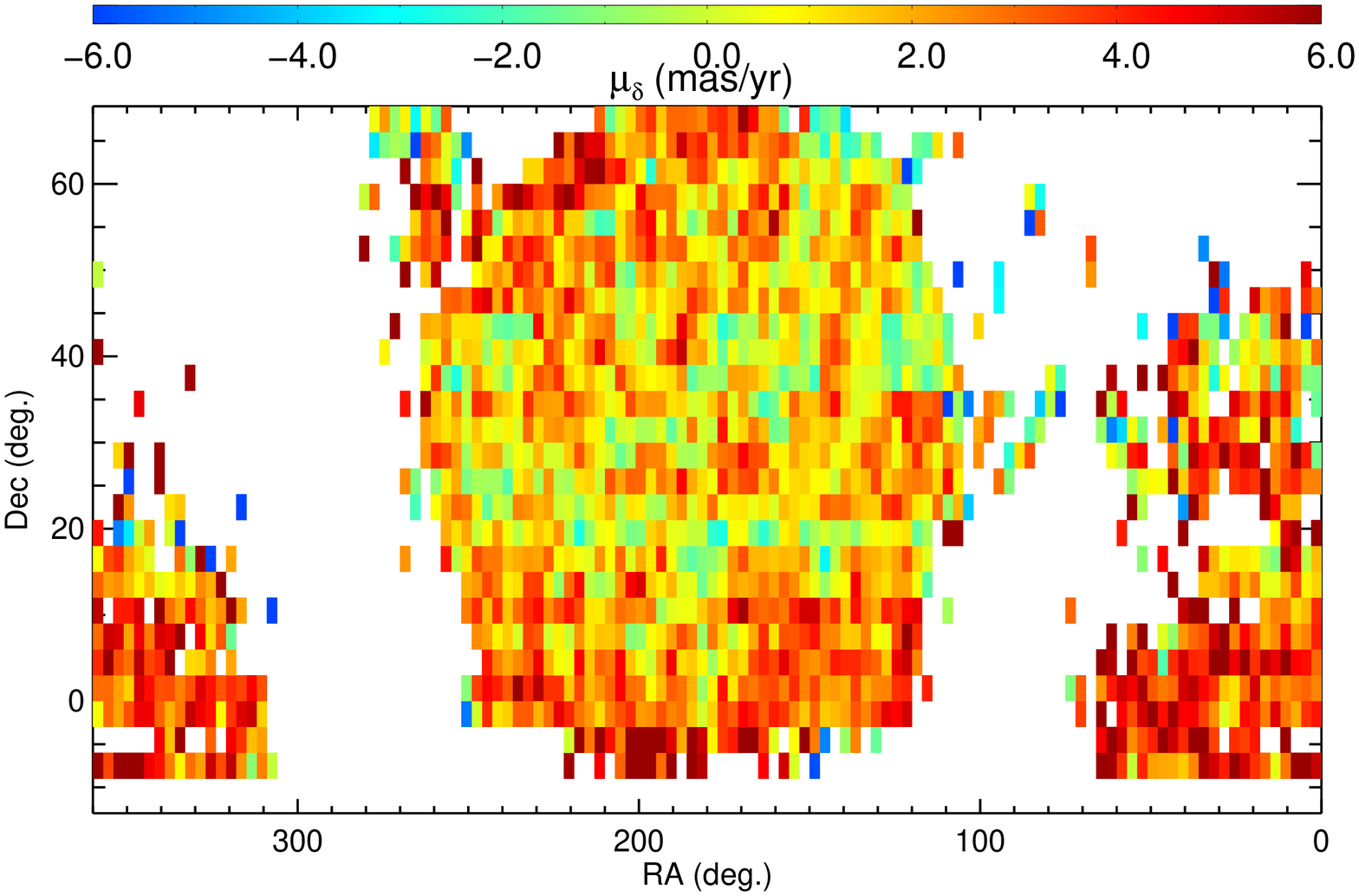}\\
\includegraphics[width=.49\linewidth]{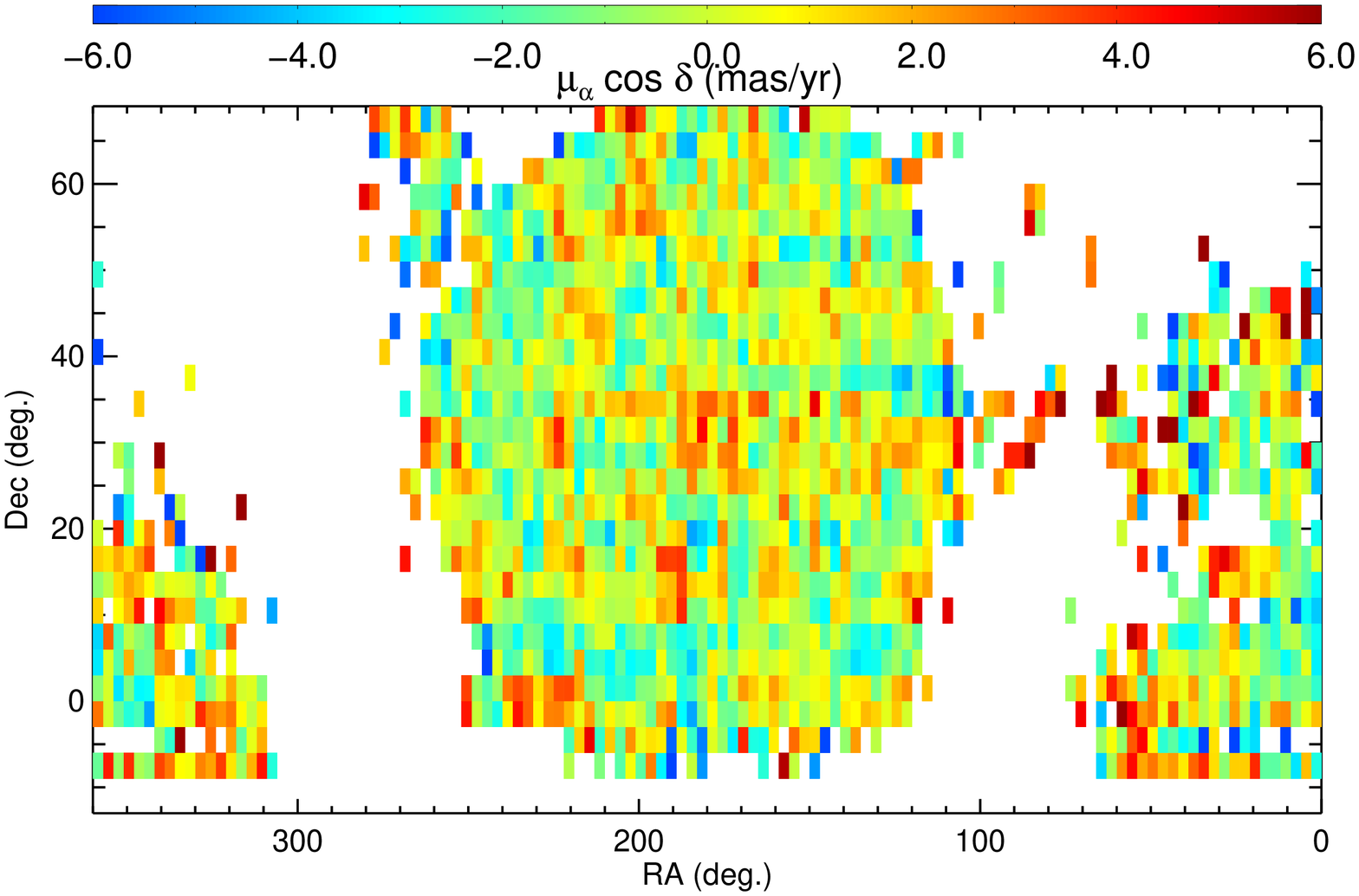}&
\includegraphics[width=.49\linewidth]{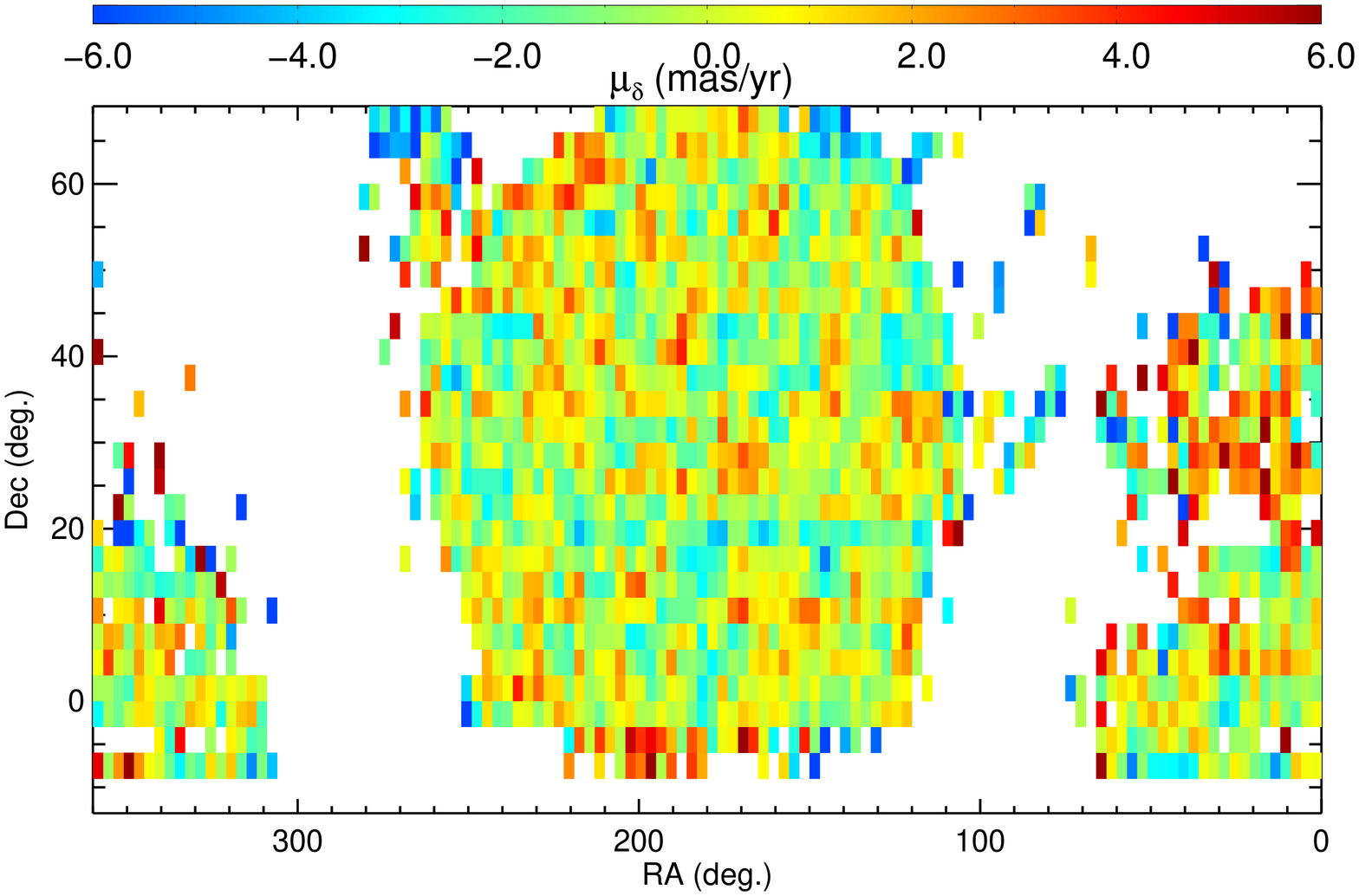}
\end{tabular}
\caption{Mean proper motions of extragalactic objects in $3\times3^\circ$ bins after correcting via the methods of \citet{wmz11} (upper panels) and \citet{cdn+13}. Color encodes the mean $\mu_\alpha \cos{\delta}$ (left panels) and $\mu_\delta$ (right panels) in mas~yr$^{-1}$. While most of the systematic error (with the exception of the remaining shift in $\mu_\delta$ in the Wu et al. panel; upper right) has been removed by each of these techniques, there are still significant variations between adjacent bins and on scales less than $\sim10^\circ$. This is the reason that we choose to simply provide look-up tables as a function of position rather than attempting to account for systematic zero-point shifts on all spatial scales. }
\label{correction}
\end{figure}

This is our motivation for suggesting the use of a look-up table of proper motion corrections in $3\times3^\circ$ areas on the sky. We find that using a sine curve to describe the trends is not effective, especially as there are not many known extragalactic objects near the Galactic plane where the sine curve of \citet{wmz11} changes from negative to positive. Polynomial corrections in two dimensions are slightly better for the large-scale behavior of proper motions in both directions. However, when we plot the corrected proper motions from these two methods on an auxiliary axis as a function of position (Figure~\ref{correction}), we still see minor variations between $3\times3$ degree bins that cannot be corrected by a smoothly-varying functional form.

\begin{table}[!t]
\centering
\begin{tabular}{ccccccccc}
\hline
RA$_{min}$ & RA$_{max}$	&	Dec$_{min}$	&Dec$_{max}$ &	$\mu_\alpha \cos{\delta}$	&	$\mu_\delta$	&	$\sigma_{\overline{\alpha}}$	&	$\sigma_{\overline{\delta}}$	&	N	\\
(deg.) & (deg.) & (deg.) & (deg.) & (mas~yr$^{-1}$) & (mas~yr$^{-1}$) & (mas~yr$^{-1}$) & (mas~yr$^{-1}$) & \\
\hline
180	&	183	&	-27	&	-24	&	1.70	&	-1.10	&	0.21	&	0.40	&	4	\\
180	&	183	&	-12	&	-9	&	-5.19	&	1.16	&	0.27	&	0.27	&	20	\\
180	&	183	&	-3	&	0	&	-3.37	&	-4.93	&	0.02	&	0.01	&	365	\\
180	&	183	&	0	&	3	&	-2.71	&	-5.78	&	0.01	&	0.01	&	494	\\
180	&	183	&	3	&	6	&	-3.86	&	-3.87	&	0.05	&	0.05	&	86	\\
180	&	183	&	6	&	9	&	-3.45	&	-5.30	&	0.07	&	0.08	&	59	\\
180	&	183	&	9	&	12	&	-3.80	&	-2.65	&	0.04	&	0.05	&	103	\\
180	&	183	&	12	&	15	&	-1.66	&	-5.39	&	0.05	&	0.06	&	87	\\
180	&	183	&	15	&	18	&	-5.53	&	-4.99	&	0.06	&	0.06	&	67	\\
180	&	183	&	18	&	21	&	-6.53	&	-3.77	&	0.05	&	0.06	&	80	\\
180	&	183	&	21	&	24	&	-4.50	&	-2.23	&	0.05	&	0.04	&	81	\\
180	&	183	&	24	&	27	&	-3.36	&	-2.10	&	0.06	&	0.07	&	85	\\
180	&	183	&	27	&	30	&	-0.16	&	-3.33	&	0.06	&	0.08	&	66	\\
\hline
\end{tabular}
\caption{A sample of the table of proper motion corrections near the anti-center of the Galaxy. For a given RA and Dec range, the sigma-clipped mean proper motion in RA and Declination, the standard error of the mean, and the number of objects in the 3x3 square degree bin used to calculate the proper motions are included. The full table is available in the electronic edition of the journal.}
\label{pmtable}
\end{table}

We present a subset of the table of proper motion corrections in Table~\ref{pmtable} (the full table is available in the electronic edition of the journal). The table lists the min/max values of the RA and Dec range for each bin, the sigma-clipped average proper motions in RA and declination for extragalactic objects, the standard error of the mean proper motions, and the number of objects in the bin used to calculate the mean values. The corrections given in this table should be subtracted from the proper motion given in PPMXL for any object within a given bin. In Figure~\ref{pms_mc} we show the proper motions of all individual extragalactic objects before (dashed histograms) and after (solid lines) the corrections. Gaussian fits to the proper motions after correcting them yield a $\sigma \sim 4$~mas~yr$^{-1}$ spread. Thus we suggest that after applying the corrections from Table~\ref{pmtable}, the random uncertainty on the zero point for an individual object is $\sim4$~mas~yr$^{-1}$, with little to no systematic offset. Because this is similar to the formal errors in PPMXL for well-measured individual proper motions, there may not be variations on scales smaller than $3^\circ$. A larger data set in the future would be useful for testing whether there are zero-point fluctuations on scales of $\sim1^\circ$.

\section{Conclusions}

We have used a large database of extragalactic sources to derive corrections to the absolute reference frame of the PPMXL proper motion catalog. The sources consist of QSOs from the V{\`e}ron Catalog of Quasars \& AGN, 13th Edition \citep{vv10}, supplemented with spectroscopically identified QSOs and galaxies from the LAMOST survey. This yields a total of 158,106 extragalactic reference objects with which to determine zero points. 

Unlike previous studies that correct the PPMXL proper motions via fits of analytical functions to the residuals of extragalactic sources \citep{cdn+13,wmz11}, we choose to construct a table as a function of position on the sky. After some experimentation with the size of bins on the sky to use, we find that $3\times3$-degree bins are a reasonable compromise that maps systematic shifts on fairly small spatial scales, while containing enough sources in each bin to robustly determine the mean residual. We derive sigma-clipped mean proper motions for the set of extragalactic objects in $3\times3^\circ$ bins (in RA and Dec) on the sky. Assuming that these extragalactic sources should have zero proper motion, the mean measured proper motions in each block can be used as corrections to place stellar proper motions from PPMXL onto an absolute frame. 

The LAMOST survey is ongoing, and will ultimately obtain spectra over much of the northern hemisphere above $\delta = -10^\circ$. In the regions already covered by SDSS, the extragalactic reference frame is already well-defined. However, near the Galactic anticenter at low ($|b| < 30^\circ$) latitudes, LAMOST will likely provide significant numbers of newly-discovered QSOs over a large contiguous area. The proper motion corrections in these low-latitude regions are extremely valuable for studies of the outer Galactic disk, among other things. Farther from the Galactic plane in the south Galactic cap, there are also few SDSS QSOs. LAMOST has already filled in many of the gaps in SDSS coverage in the southern Galactic sky, and will ultimately provide a huge database with which to improve the proper motion systematics in this region. As the LAMOST survey continues, the area of sky observed will grow, and the additional sources identified in regions previously observed will improve the statistics in bins presented here. We expect that at the completion of the LAMOST survey, we will present a table of proper motion zero-point corrections for a large fraction of the northern sky.

\section{Acknowledgements}

This research is supported by the National Science Foundation under Grant AST 09-37523. T.C.B. acknowledges partial support for this work from grant PHY 08-22648; Physics Frontier Center/Joint Institute for Nuclear Astrophysics (JINA), awarded by the US National Science Foundation. Guoshoujing Telescope (the Large Sky Area Multi-Object Fiber Spectroscopic Telescope, LAMOST) is a National Major Scientific Project built by the Chinese Academy of Sciences. Funding for the project has been provided by the National Development and Reform Commission. LAMOST is operated and managed by the National Astronomical Observatories, Chinese Academy of Sciences.

\label{lastpage}

\end{document}